# Ensemble Classification-Based Spectrum Sensing Using Support Vector Machine for CRN

[1]Manpreet Kaur, [2]Raj Singh [3]Sandeep Kumar*

[1,2,3] Central Research Laboratory, BEL, Ghaziabad, India

Corresponding Author Mailid: sann.kaushik@gmail.com

*Abstract—* **As the demand for internet of things (IoT) and device-to-device (D2D) applications in next generation communication systems increases, we are confronted with a challenge of spectrum scarcity. One promising solution to this problem is cognitive radio network (CRN), where the key element is the spectrum - a valuable and sharable natural resource that should not be wasted. To design efficient and sustainable networks for the future, it is crucial to ensure that spectrum sensing is not only accurate and rapid, but also energy-efficient. Spectrum sensing is a critical aspect of CRNs, and this study is mainly focused on it. In this research, we employ the supervised machine learning algorithm, support vector machine (SVM), to detect primary users (PU). We investigate different variants of SVM, including linear, polynomial, and Gaussian radial basic function (RBF), and employ an ensemble classification-based approach to improve the classifier's performance and productivity. The simulation results demonstrate that the ensemble classifier achieves the highest performance.**

1. INTRODUCTION

The growing usage of wireless traffic networks, driven by the increasing demand for internet of things (IoT) and device-to-device (D2D) communication applications in future 6th generation (6G) wireless networks, has resulted in a severe shortage of radio spectrum worldwide. To address this issue and provide bandwidth to emerging applications through dynamic spectrum access (DSA), cognitive radio network (CRN) technology has emerged as a promising solution. In the CRN, there are two types of users: licensed user or primary user (PU) and unlicensed users or secondary user (SU) [2]. SUs can only access the licensed spectrum when it is not being used by the corresponding PU, thus effectively reducing the spectrum shortage problem. In the CR network, SUs must be careful to ensure that they do not cause any interference to the PU while accessing the PU's spectrum. Therefore, spectrum sensing is the main function of SUs, and they are equipped with software-defined radio (SDR) that allows them to reconfigure their network parameters based on the network environment [3].

In the literature, energy detection is the most commonly used approach for spectrum sensing (SS) due to its simple circuitry and low complexity [2]. However, the detection of the spectrum for a single node SU is prone to error due to the random nature of fading channels between the primary user PU and the SU. To address this, collaborative spectrum sensing (CSS) has been proposed, where multiple SUs perform individual SS and combine their results to make a final decision, thus improving the SS performance of the system. However, designing a generic CSS strategy for different channel conditions is challenging, as the scenarios for SUs that are close to the PU differ from those that are far away. The performance of CSS using the energy detector (ED) over $\alpha-\eta-\mu$ fading channel has been analyzed in terms of receiver operating characteristic (ROC) and area under the receiver operating characteristic curve (AUC) in [2]. In [3], the ED performance over $\alpha-\eta-\mu$/IG and $\alpha-\kappa-\mu$/IG is evaluated, while [4] studies the detection performance over $\alpha-\eta-\mu$/gamma and $\alpha-\kappa-\mu$/gamma composite fading channels. The work in [4] and [5] is extended for CSS, and the performance improvement of the system using a cooperative approach has been demonstrated. The most popular CSS strategies used in the literature are ORing, ANDing, and K-out-of-N, where the individual sensing of SUs is considered to be independent of each other. Although the above CSS approaches have shown performance improvement over the individual sensing scheme, their performance may degrade in the dynamic network environment.

Regarding spectrum sensing in cognitive wireless radio networks, machine learning (ML) algorithms are suitable for detecting the presence of the PU signal. As ML techniques can recognize patterns in data, they can be used to classify a set of data into different categories, such as detecting the presence or absence of the PU signal. Many ML algorithms have been proposed for spectrum sensing in cognitive wireless radio networks, and a survey of these methods is provided in [6].

An unsupervised learning approach utilizing K-means clustering has been employed to detect white spaces in the spectrum using a real-time, hardware-implementable spectrum sensor that has been developed and tested, as described in reference [7]. Another approach presented in reference [8] utilizes a combination of the Null Space Pursuit algorithm (NSP) and fuzzy c-means (FCM) clustering algorithm to sense the signal using pre-processing, decomposing the signal into sub-signal components that exhibit more distinct features. In reference [9], a CSS method based on information geometry and fuzzy c-means clustering algorithm is introduced. Finally, reference [10] describes a spectrum sensing method for Cognitive Wireless Multimedia Sensor Networks that utilizes clustering algorithms and signal features.

The α–κ–μ fading channel is a generalized fading channel model that can encompass many other fading channel models [11]. Despite this, there is currently no literature on the performance analysis of ML-based CSS over the α–κ–μ generalized fading channel. To address this gap, we propose an ensemble classifier based on support vector machines (SVM) to accurately detect the presence of a PU signal. Our contributions in this paper can be summarized as follows:

- We develop an ensemble classifier that combines SVM with different kernel functions, including linear, polynomial, and Gaussian radial basis function (RBF) kernels, to improve the accuracy of spectrum occupancy assessment.
- We demonstrate that the proposed scheme outperforms all individual SVM schemes in the wireless environment scenario considered in this study.
- The α–κ–μ fading channel model used in this study is a generalized fading channel that can subsume many other fading channel models [11], meaning that our results can be directly applied to special cases of the α–κ–μ channel by adjusting the fading parameters.

The paper is organized as follows. Section 2 describes the system model, while Section 3 presents the proposed framework. In Section 4, we evaluate the performance of the system using various performance parameters and provide numerical results. Finally, we conclude the paper in Section 5.

## 2. SYSTEM MODEL

### 2.1 Cognitive Radio Network

In a CRN, the SUs are allowed to use the frequency band of PUs, when it is not in use by the PU. However, if an SU, such as SUa, is currently using the frequency band of a PU, such as PUa, and PUa suddenly wants to use their band, SUa must vacate the band and switch to another available band. Therefore, the SU must continuously sense the available spectrum in order to take advantage of the unlicensed service offered by the network [12]. However, spectrum sensing becomes challenging for the SU due to the distance between the PU and SU, which can lead to small-scale and large-scale fading that negatively affects signal propagation [13]. In a CRN, SUs are required to use the PU spectrum without interfering with PU operation. Thus, an incorrect decision by the SU about the presence of a PU signal would violate this condition.

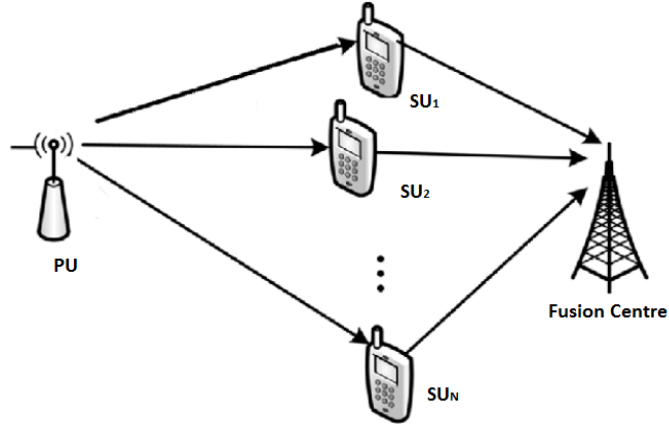

Figure 1 CRN with cooperative spectrum sensing

In order to enhance the ability of sensing and decision-making regarding the presence of the PU signal, researchers have turned to cooperative methods, as documented in the literature [14]. Figure 1 depicts a cognitive radio scenario in which the PU signal travels to SUs after undergoing various propagation phenomena. The signal samples from the PU's band are received by all SUs.

2.2 Energy Detector

The spectrum sensing technique known as energy detector is commonly used due to its low computational complexity and lack of need for prior knowledge of the PU signal. By measuring the signal power on the target frequency bands and comparing the results to a predefined threshold, the spectrum occupancy of the PU can be determined. This method is well suited for 5G networks where prompt and timely detection of spectrum availability is necessary [15]. The energy of the received signal is utilized as a feature vector for spectrum sensing, with the normalized energy estimated at the nth SU as

$$Y_n = \frac{1}{M} \sum_{i=1}^{M} |Z_n(i)|^2 \qquad (1)$$

where $M$ is the number of samples and $Z_n(i)$ is the $i^{th}$ sample of the received signal at the $n^{th}$ SU.

$$Z(i) = \begin{cases} w(i); & H_0 \\ h*s(i) + w(i); & H_1 \end{cases} \qquad (2)$$

where $w(i)$, $s(i)$ and $h$ are additive white Gaussian noise (AWGN), the sample of the transmitted symbol and channel gain respectively.

## 2.3 Channel Model

The α–κ–μ fading distribution is a generalized model used to represent real-world signal propagation scenarios, including line-of-sight (LOS) conditions [5]. Many other multipath fading models, such as Rayleigh, Ricean, Nakagami, Weibull, α–μ, and κ–μ, are special cases of this generalized distribution. The probability density function (PDF) of the received signal-to-noise ratio (SNR) over an α–κ–μ fading channel is provided by the following equation

$$f_\gamma(\gamma) = \frac{\mu \alpha k^{\left(\frac{1-\mu}{2}\right)} (1+k)^{\left(\frac{1+\mu}{2}\right)} \left(\sqrt{\frac{\gamma}{\overline{\gamma}}}\right)^{\left(\frac{\alpha}{2}+\frac{\alpha\mu}{2}-1\right)}}{2\sqrt{\gamma\overline{\gamma}} \exp\left[\mu\left(k+(1+k)\left(\sqrt{\frac{\gamma}{\overline{\gamma}}}\right)^\alpha\right)\right]}$$
$$\times I_{(\mu-1)}\left(2\mu\sqrt{k(k+1)\left(\sqrt{\frac{\gamma}{\overline{\gamma}}}\right)^\alpha}\right) \tag{3}$$

The parameters α and μ are utilized to describe the non-uniformity and non-linearity of the propagation channel, while κ is the ratio of the in-phase and quadrature components' scattered-wave power [5].

$$f_Y(y) = \begin{cases} \dfrac{1}{2^n \Gamma(n)} y^{n-1} \exp\left(\dfrac{-y}{2}\right) & ; H_0 \\ \dfrac{1}{2}\left(\dfrac{y}{2\gamma}\right)^{n/2-1/2} \exp\left(\dfrac{-y-2\gamma}{2}\right) I_{n-1}\left(\sqrt{2y\gamma}\right) & ; H_1 \end{cases} \tag{4}$$

The term $n$ is the product of time and bandwidth and should be an integer representing the number of time samples. T represents the integration time while W denotes the bandwidth of the system. The modified Bessel function is represented as $I_v(.)$ and while the gamma function is denoted by $\Gamma(.)$. The received SNR is given by $\gamma = |h|^2 E_S^2/N_0$ where $E_S$ represents the signal energy and $N_0$ is the one-sided power spectral density (PSD) [10]. In ED-based CSS, each of the N SU nodes shares their decisions with the fusion center (FC) to arrive at a rule-based decision about the PU signal.

## 3. PROPOSED FRAMEWORK

The proposed approach involves having each SU perform spectrum sensing locally using energy detection technique. The individual sensing results are then transmitted to a FC, which constructs energy level vectors for classification. To counteract the fading effects experienced by individual SUs, cooperative decision-making at the FC is employed, leveraging spatial and diversity gains obtained from the collaborating SUs. Prior to online classification, the proposed method undergoes a training phase, during which a significant number of training samples are collected. The trained system then classifies subsequent energy levels from the cooperating SUs into separate classes to make decisions on spectrum occupancy.

3.1 Support Vector Machine

SVM is a supervised learning technique that utilizes classification algorithms for problems with two groups. In SVM, the energy feature vectors for training must be accompanied by a label indicating the corresponding class. After obtaining the training energy vectors, different kernel functions are used for learning purposes. These functions map input training energy vectors into high-dimensional feature spaces to linearly separate classes. This work utilizes three kernel functions: linear, polynomial, and Gaussian radial basis function (RBF).

Let $\{(D_1,l_1),(D_2,l_2)...,(D_L,l_L)\}$ represent the training data and the corresponding labels of the L energy vectors of individual SUs participating in the cooperative decision-making. The label $l=1$ indicates the presence of the PU (represented by blue diamond in Figure 2) while $l=-1$ denotes the absence of any PU (represented by green circle in Figure 2) in the received signal. The objective of SVM is to find the optimal hyperplane that separates the positive and negative classes, denoted by $2/\|w\|$ as shown in Figure 2.

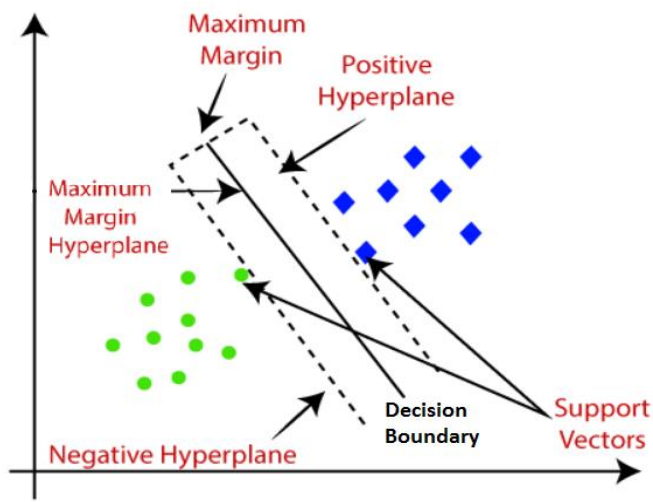

Figure 2 SVM Model

Under severe fading conditions, linearly separating the training samples is challenging [16]. To overcome this challenge, a non-linear kernel function $\phi(D)$ is introduced to map the training data into higher-dimensional feature space in order to linearly separate the classes. Therefore, the classifier is expected to satisfy the following conditions:

$$\langle w.\phi(D_i)\rangle + \beta \geq 1 \quad \forall \ Y_i = 1$$
$$\langle w.\phi(D_i)\rangle + \beta \leq -1 \quad \forall \ Y_i = -1 \tag{5}$$

where $w$ is the weighing vector, $\beta$ is the bias that moves the hyperplane away from the origin and the quantity $\langle . \rangle$ is the inner product. Equation (5) can be rewritten as follows:

$$Y_i\left(\langle w.\phi(D_i)\rangle + \beta\right) \geq 1 \quad \forall i \tag{6}$$

It is not always possible to achieve an ideal, linearly separable hyperplane that satisfies (6) for each training data set. To prevent overfitting of the data, a slack variable $\rho$ is introduced for potential classification errors. Therefore, the above equation becomes:

$$Y_i\left(\langle w.\phi(D_i)\rangle + \beta\right) \geq 1 - \rho_i \quad \forall i, \rho \geq 0 \tag{7}$$

The training samples for which $\rho = 0$ are considered to be classified correctly and those that lie inside $0 \leq \rho_i \leq 1$ are considered to be within marginal classification errors and are on the correct side of the decision boundary. Conversely, if $\rho_i \geq 1$, it is considered a classification error. The goal is to minimize the sum error while maximizing the classifier's margin. This can be achieved as follows:

$$\min \frac{1}{2}\|w\|^2 + \theta \sum_{i=1}^{S} \rho_i \ \text{Subject to} \ Y_i\left(\langle w.\phi(D_i)\rangle + \beta\right) \geq 1 - \rho_i \quad \forall i, \rho \geq 0 \tag{8}$$

where $\|w\|^2$ is the inner product given by $w^T.w$ while $\theta$ is a soft margin constant [17]. The resulting optimization problem can be solved using Lagrangian function as

$$L(w,\beta,\rho,\alpha,\delta) = \begin{cases} \min \frac{1}{2}\|w\|^2 + \theta \sum_{i=1}^{P} \rho_i \\ -\sum_{i=1}^{S} \alpha_i\left[Y_i\left(\langle w.\phi(D_i)\rangle + \beta\right) - 1 - \rho_i\right] \\ -\sum_{i=1}^{S} \delta_i \rho_i \end{cases} \tag{9}$$

where $\alpha_i$ and $\delta_i$ are Lagrangian multipliers. The training samples where $\alpha_i \geq 0$ are referred to as support vectors, and lies on one of the two hyperplanes. Applying the Karush-Kuhn-Tucker (KKT) condition to (9), we obtain:

$$w = \sum_{i=1}^{S} \alpha_i Y_i \phi(D_i)$$
$$\sum_{i=1}^{S} \alpha_i Y_i \phi(D_i) \quad (10)$$
$$\alpha_i + \delta_i = \theta$$

Since $\alpha_i \geq 0$, we have $0 \leq \alpha_i \leq \theta$ with $\theta$ setting the upper bound. In addition, since the values of $\alpha_i$ are support vectors, the dual form of the problem defined in (10) in terms of support vectors can be formulated as

$$L(w,\beta,\rho,\alpha,\delta) = \left[ \sum_{i=1}^{P} \alpha_i - \sum_{i=1}^{P}\sum_{j=1}^{P} Y_i Y_j \alpha_i \alpha_j \langle \phi(D_i)\phi(D_j) \rangle \right]$$
$$\text{Subject to } \sum_{i=1}^{P} Y_i \alpha_i = 0, \; 0 \leq \alpha_i \leq \theta \quad \forall i \quad (9)$$

By solving the convex optimization problem in the above equation and applying the quadratic programming algorithm, the non-linear decision function is obtained as follows:

$$Y(X) = \text{sgn}\left( \sum_{i=1}^{S} \alpha_i Y_i k(X, X_i) + \beta \right) \quad (10)$$

where $sgn$ is the sign function and $k(X, X_i) = \langle \phi(D_i)\phi(D_j) \rangle$ is the kernel function. Different types of kernel function can be used, including linear, polynomial and Gaussian radial basis function [17]. The linear kernel function is given by: $k(X, X_i) = X_i^T . X_j$, polynomial kernel: $k(X, X_i) = (X_i^T . X_j + 1)^d, d > 1$ and the radial basic function (RBF) kernel: $k(X, X_i) = \exp\left( \frac{\|X_i - X_j\|}{2\sigma^2} \right)^T$ [17].

3.2 Ensemble classifier

Ensemble learning involves creating multiple base classifiers that can be combined to form a new classifier which outperforms any individual classifier. The base classifiers can differ in their algorithm, hyper parameters, representation, or training set. The main goal of ensemble methods is to minimize bias and variance.

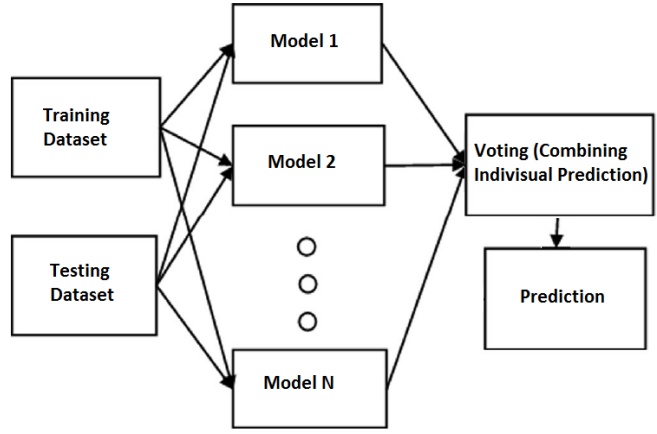

Figure 3 Ensemble Classifier

To conduct a comprehensive comparison between the ML approaches used in this case study, we employ an ensemble classifier, as depicted in figure 3. This ensemble classifier consists of various SVM variants as individual classifier models and an algorithm to combine them. To achieve this, we adopt stacking, which involves training multiple independent classifiers, each trained by sampling a certain percentage of instances from the training data with replacement. The diversity in the ensemble is ensured by the variations in replicas on which each classifier is trained, which enhances the classifier's predictive power and performance.

4. PERFORMANCE PARAMETERS

To evaluate the performance of the proposed scheme, we use ROC curves and compare the simulation results to the hard/soft fusion rule at the FC. The ROC curve is a graphical tool that illustrates the relationship between the probability of detection and the probability of false alarm, which reflects the performance of a classification model at different classification thresholds.

The performance parameters are defined by [18] as

$$P_{fa} = P\left[A^* = -1 \mid A = +1\right]$$
$$P_d = P\left[A^* = -1 \mid A = -1\right]$$
(11)

where A is the actual state and A* is the predicted state of the channel.

5. NUMERICAL RESULTS.

The proposed spectrum sensing-based ML models using four classifiers are evaluated through different performance metrics, including AUC and ROC curves. This section discusses the impact of system parameters on the detection performance of the proposed model.

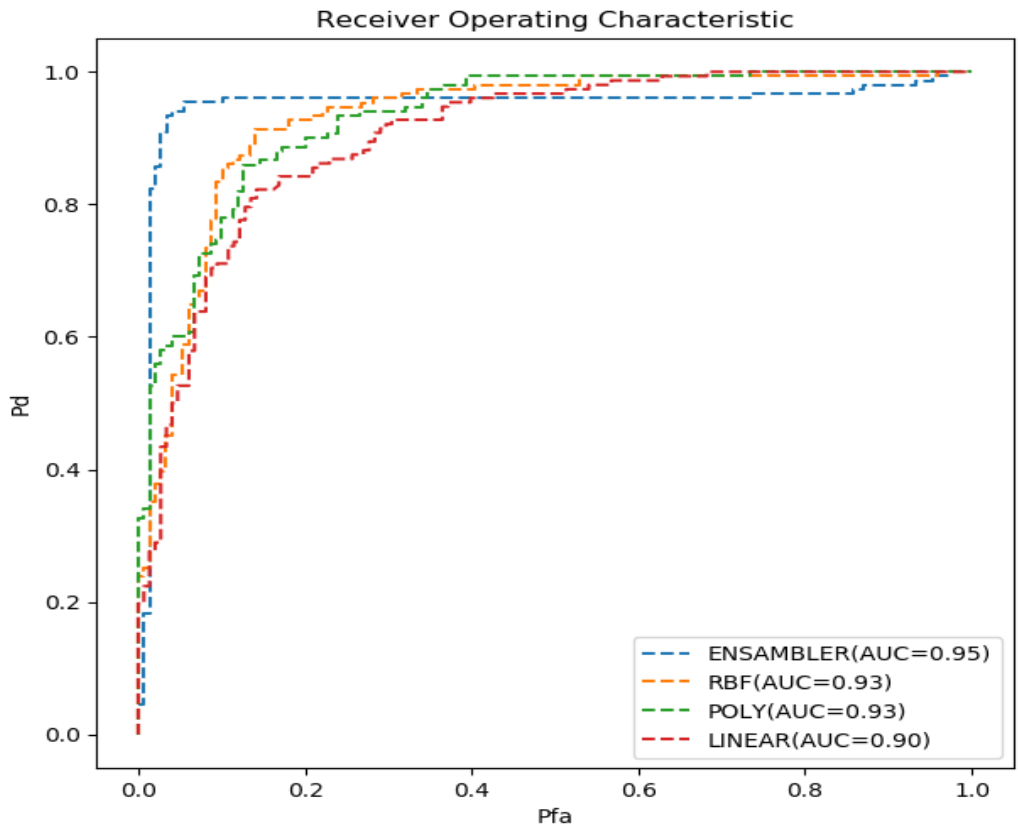

Figure 4 Comparison of different classifiers

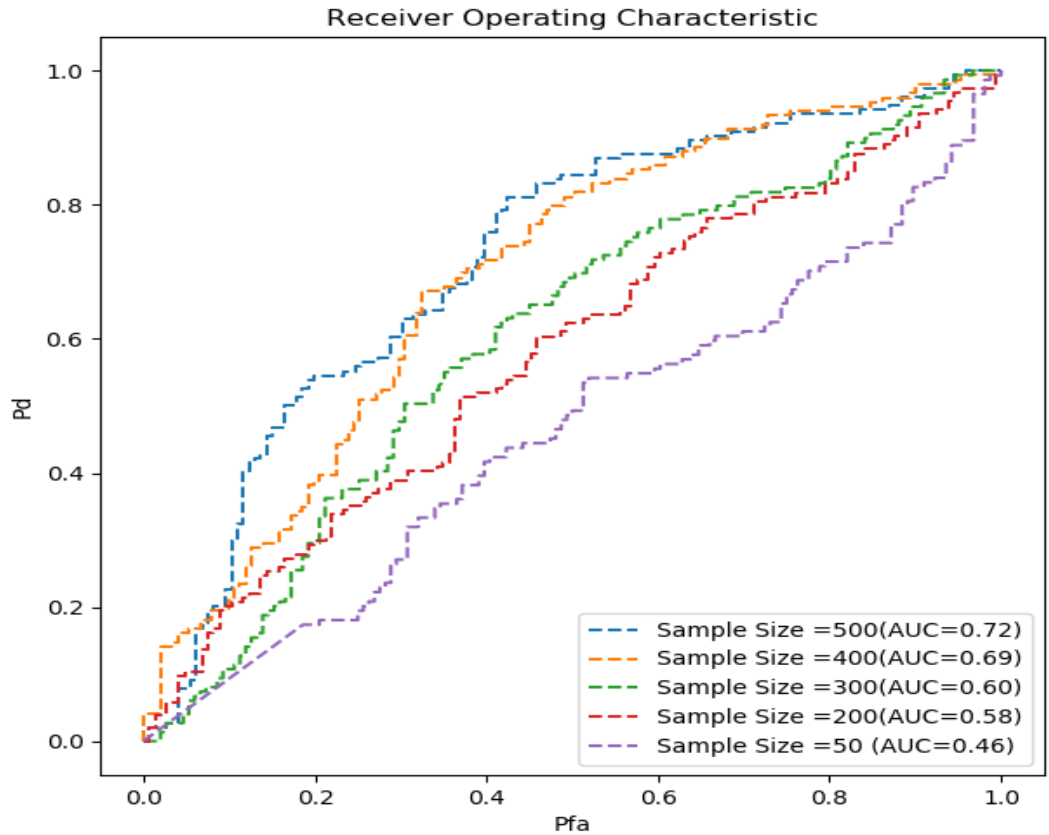

Figure 5 The impact of sample size on the ROC

Figure 4 displays the ROC curves of the four classifiers, including Linear, Polynomial, RBF, and Ensemble classifiers of SVM classifiers. The Ensemble classifier outperforms the individual SVM classifiers, since it combines multiple classifiers, while the RBF SVM classifier comes in second place by finding the hyper-plane that maximizes the margin between the classes. The system's sample size impact on the ROC and AUC performance is illustrated in Figure 5, showing that the system performance increases as the sample size increases. Figure 6 shows the ROC curve of the proposed ensemble algorithm plotted for various average SNR values. The graph clearly demonstrates the positive influence of a high SNR on the performance of both the ROC and AUC metrics. Figure 7 illustrates the improved performance of the ROC plots with a high number of SUs. Although the ROC plots cross each other at certain points, the superior performance for a high number of SUs is evident from the AUC values. However, one of the drawbacks of the SVM-based approach is that it necessitates training before online sensing can be performed, and online detection is restricted to the particular channels on which the model was trained. Nonetheless, as more data is gathered on numerous channels, the sensing performance for many channels can be obtained and improved upon during the learning process.

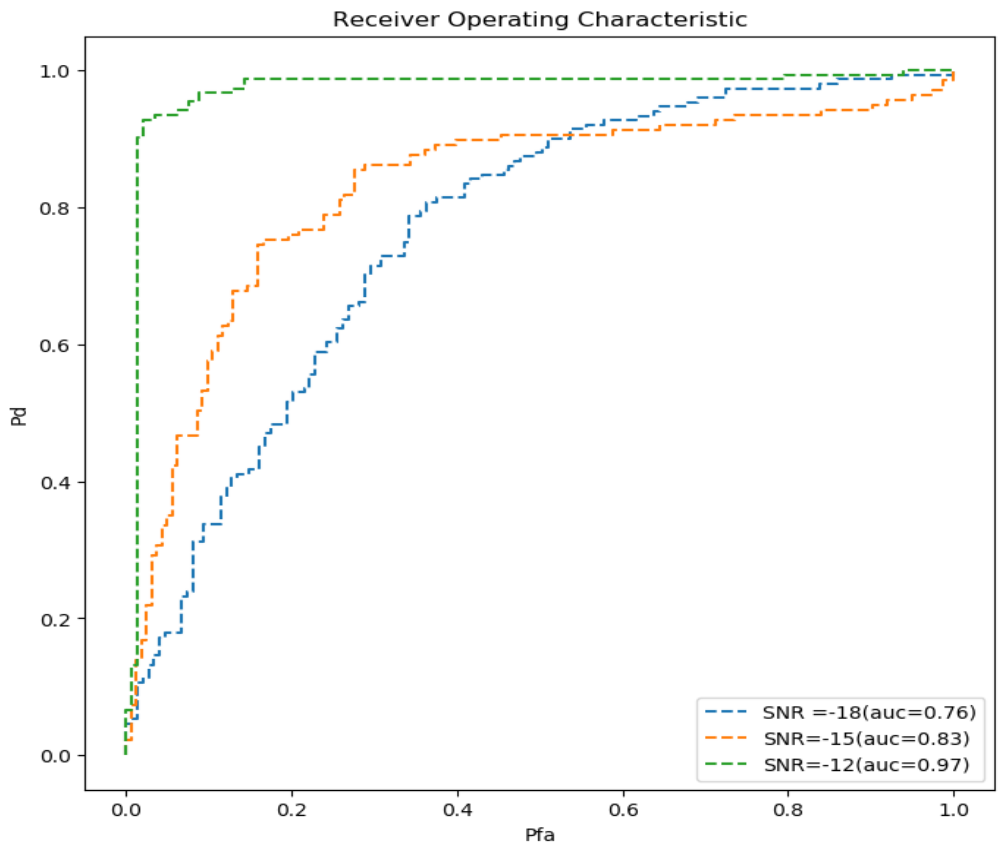

Figure 6 Pd versus Pfa for various values of average received SNR

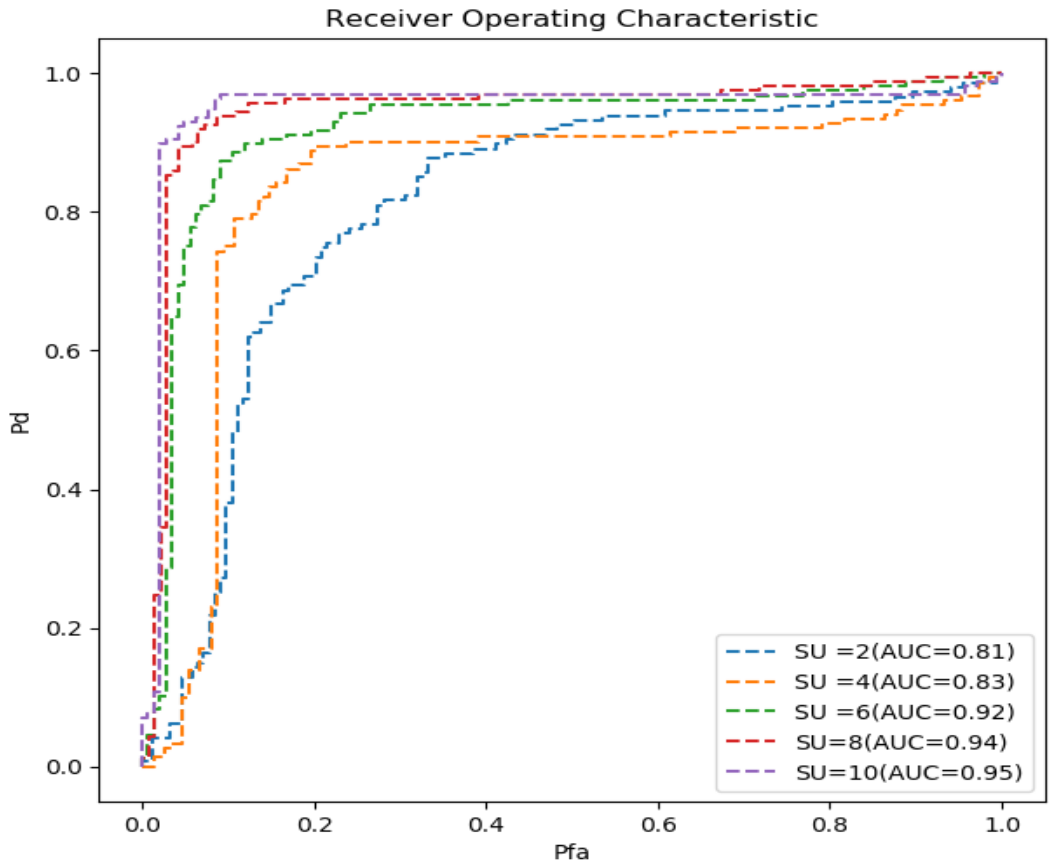

Figure 7 Comparison of the ROC curves for different number of SUs

## 6. CONCLUSIONS

In this study, the detection of PUs using the SVM was investigated. Different variants of SVM, including linear, polynomial, and RBF, were explored. In addition, an ensemble classification-based approach was employed to improve classifier productivity and performance. Simulation results demonstrated that the ensemble classifier achieved the best performance, followed by the RBF SVM classifier, based on performance metrics such as AUC and ROC. However, this improved performance comes at the cost of increased computational complexity. The results indicate that the system's performance improves with an increase in the number of SUs contributing to decision-making, the received SNR, and the sample size.

**Declarations:**

**Code availability** N/A

**Conflicts of Interest** There is no conflict of interest.

**Authors' contributions** All the authors have equally contributed in this manuscript.

**Data Availability** N/A (There is no research data outside the submitted manuscript file.)

**Funding** No funding was received.